\documentclass[%
reprint,
superscriptaddress,
amsmath,amssymb,
aps,
floatfix,
]{revtex4-2}
\usepackage{graphicx}
\usepackage{subcaption}
\usepackage{siunitx}
\usepackage{csquotes}
\usepackage{bm}
\usepackage{physics}
\usepackage{mathtools}
\usepackage{amsfonts}
\usepackage{amssymb}
\usepackage{amsmath}
\usepackage{diagbox}
\usepackage{hyperref}
\usepackage{xspace}
\hypersetup{linktocpage,colorlinks,citecolor={blue},pdfdisplaydoctitle=true,pdfpagemode=UseOutlines,bookmarksnumbered=true}
\usepackage{dsfont}
\usepackage{todonotes} 
\usepackage{booktabs}
\usepackage[most]{tcolorbox}

\usepackage{afterpage}

\usepackage[T1]{fontenc}

\mathchardef\mhyphen="2D 

\usepackage{breakurl}
\usepackage{url}

\usepackage{bbold}

\usepackage{siunitx}

\newcommand{\aqa}{$\langle aQa ^L\rangle $ Applied Quantum Algorithms, Universiteit Leiden}
\newcommand{\lorentz}{Instituut-Lorentz, Universiteit Leiden, Niels Bohrweg 2, 2333 CA Leiden, Netherlands}
\newcommand{\liacs}{LIACS, Universiteit Leiden, Niels Bohrweg 1, 2333 CA Leiden, Netherlands}
\newcommand{\amherst}{Manning College of Information and Computer Sciences, University of Massachusetts Amherst, 140 Governors Dr, Amherst, MA 01002, United States}
\newcommand{\delft}{EEMCS, Delft University of Technology, Mekelweg 4, 2628 CD Delft, The Netherlands}
\newcommand{\qutech}{QuTech, Delft University of Technology, Lorentzweg 1, 2628 CJ Delft, The Netherlands}

\begin{document}
\title{Optimising entanglement distribution policies under classical communication constraints assisted by reinforcement learning}
\author{Jan Li}
\affiliation{\aqa}
\affiliation{\lorentz}

\author{Tim Coopmans}
\affiliation{\aqa}
\affiliation{\liacs}
\affiliation{\qutech}
\affiliation{\delft}

\author{Patrick Emonts}
\affiliation{\aqa}
\affiliation{\lorentz}

\author{Kenneth Goodenough}
\affiliation{\amherst}

\author{Jordi Tura}
\affiliation{\aqa}
\affiliation{\lorentz}

\author{Evert van Nieuwenburg}
\affiliation{\aqa}
\affiliation{\liacs}

\date{\today}

\begin{abstract}
Quantum repeaters play a crucial role in the effective distribution of entanglement over long distances. 
The nearest-future type of quantum repeater requires two operations: entanglement generation across neighbouring repeaters and entanglement swapping to promote short-range entanglement to long-range.
For many hardware setups, these actions are probabilistic, leading to longer distribution times and incurred errors.
Significant efforts have been vested in finding the optimal entanglement-distribution policy, i.e. the protocol specifying when a network node needs to generate or swap entanglement, such that the expected time to distribute long-distance entanglement is minimal.
This problem is even more intricate in more realistic scenarios, especially when classical communication delays are taken into account.
In this work, we formulate our problem as a Markov decision problem and use reinforcement learning (RL) to optimise over centralised strategies, where one designated node instructs other nodes which actions to perform.
Contrary to most RL models, ours can be readily interpreted.
Additionally, we introduce and evaluate a fixed local policy, the `predictive swap-asap' policy, where nodes only coordinate with nearest neighbours.
Compared to the straightforward generalization of the common swap-asap policy to the scenario with classical communication effects, the `wait-for-broadcast swap-asap' policy, both of the aforementioned entanglement-delivery policies are faster at high success probabilities.
Our work showcases the merit of considering policies acting with incomplete information in the realistic case when classical communication effects are significant.
\end{abstract}

\maketitle

\section{Introduction}
The Quantum Internet promises to enable a wide range of tasks that are more efficient than their classical counterparts~\cite{wehner2018quantum}. 
Examples include generating keys for informationally secure communication~\cite{bennett_quantum_2014,ekert_quantum_1991}, universal blind quantum computing~\cite{broadbent_universal_2009}, implementing interferometric telescopes with arbitrary long baselines~\cite{gottesman_longer-baseline_2012}, and improved clock synchronisation~\cite{komar_quantum_2014}. 

Crucial to the realisation of a quantum internet is the ability to entangle distant quantum systems with each other. 
Remote entanglement could in principle be generated by locally entangling a quantum memory and a photon, sending the photon to the other party, after which a local operation on the photon and a remote party's quantum memory results in remote entanglement~\cite{munro_inside_2015}.
Unfortunately, photon loss in the transmission medium renders this approach intractable for large distances due to rapidly increasing loss probability as function of the distance~\cite{munro_inside_2015}.
If the signals were classical, we could use amplification techniques, but due to the no-cloning theorem~\cite{wootters_single_1982}, this is not possible in the quantum setting. 
However, \emph{quantum repeaters} offer a solution~\cite{briegel_quantum_1998, munro_inside_2015, sangouard_quantum_2011, azuma_quantum_2023}.

Quantum repeaters are intermediate stations placed between distant systems.
Through first dividing up the full length into smaller segments, generating entanglement on those, and then later connecting these together again through entanglement swaps, entanglement over the full length can be established~\cite{bennett_teleporting_1993, zukowski_event-ready-detectors_1993, azuma_quantum_2023}.
For the first generation of quantum repeaters, the hardware will only allow for probabilistic generation and swapping of entanglement~\cite{munro_inside_2015}.
In case the action fails, the involved entanglement is lost.
Therefore, actions may need to be repeated multiple times in order for an end-to-end link to be established.
Nonetheless, performing the swapping and entanglement generation in a specific order, combined with storing of entanglement in quantum memories, can bring down the scaling of the average entanglement-delivery time from exponential to polynomial in the distance~\cite{briegel_quantum_1998,munro_inside_2015}.

Determining the optimal policy---i.e., which actions to take for a given entanglement configuration in a quantum network---to minimise the expected entanglement delivery time for specific hardware parameters is an active area of research.
Previous work has focused on analysing policies, see \cite{goodenough_noise_2024, coopmans_improved_2022, kamin_exact_2023, dai_entanglement_2021, guedes2024analysis}
as well as \cite{azuma_quantum_2023} and references therein, and recently also on automated policy optimisation~\cite{shchukin_optimal_2022, inesta_optimal_2023,haldar_fast_2024, silva_optimizing_2021}.

Most existing work assumes that classical communication between repeater nodes is instantaneous and that the nodes have a complete picture of the entanglement which is present in the repeater chain.
The only means to obtain this picture is by waiting for information from other nodes on the actions they have performed in the past --- entanglement generation or swapping --- and whether these actions were successful.
Consequently, when implemented in the real world, the nodes would spend a considerable amount of time waiting for information.
This not only increases the time until the final long-distance entanglement is delivered, but also impacts its quality if the quantum memories in which intermediate entanglement is stored are imperfect.
Indeed, in reality, communication time is not negligible, consisting of the transmission speed of classical information -- which is theoretically limited by the speed of light -- and the time overhead by the classical control and communication hardware~\cite{delledonne2024design}.

In this work, we relax this limitation and develop entanglement distribution policies based on reinforcement learning that take classical communication delays into account.
In each time step, the reinforcement learning agent, which is located at a designated node, instructs the nodes of the network to wait, to generate entanglement, to swap, or a combination of those.
It bases its instructions on past actions and the corresponding success/fail messages received from the nodes.
We compare the results obtained through reinforcement learning with different, fixed policies, duly adapted from the literature.
These fixed policies are constructed by modifying the swap-asap policy, which is optimal in the ideal scenario where the entanglement swap is deterministic and entanglement is never discarded~\cite{kamin_exact_2023, goodenough_noise_2024}.
They are adapted to either wait for the delayed results to return (wait-for-broadcast swap-asap) or try to guess the result of the actions based on the success probabilities, in order to act faster (predictive swap-asap).
When success probabilities are high, the predictive swap-asap policy performs considerably better, and becomes optimal for unit success probabilities.
The reinforcement learning policy manages to beat these strategies in the intermediate regime, where probabilities are high, but not identical to one.

Our approach is similar to the ones found in~\cite{haldar_fast_2024,inesta_optimal_2023, reis_deep_2023} where reinforcement learning techniques were used to optimise protocols without classical communication delays. 
However, in these works the entire state of the quantum network was used as an observation for the machine learning agent to learn and base its decisions on. 
With classical communication delays, complete information about the quantum network state is generally not available.
In our work, we choose a history-based approach to formulate the delayed problem as a Markov decision process.

This paper is organised as follows. 
In Section~\ref{repeater chains}, we introduce linear repeater chains, a specific type of quantum network and in Section~\ref{previous works}, we provide an overview of related previous works.
Subsequently, we explain the influence the effects of classical communication on these policies in section~\ref{CC_delays}. 
In Section~\ref{MDP formulation}, we will specify how we phrase the problem of optimising policies with classical communication delays as a Markov decision process and our approach to solve it. 
Then, we introduce the modified swap-asap policies with classical communication effects in Section~\ref{fixed_policies} and state state the reinforcement learning algorithm for the policy optimization in Section~\ref{RL_algorithm}. 
In Section~\ref{NumRes}, we present the results of our numerical optimization and in Section~\ref{conclusion outlook} we give our conclusion and outlook.

\section{Background} \label{background}

\subsection{Linear repeater networks} \label{repeater chains}

A repeater network is a collection of nodes (each of which contains some number of qubits) with limited connectivity between the them.
 In this work, we focus on networks arranged in a one-dimensional geometry, also known as quantum repeater chains~\cite{briegel_quantum_1998}.
We label the $n$ nodes in such a chain by integers $i \in [n] :=\{0,...,n-1\}$.
To establish end-to-end links, i.e. between node $0$ and $n-1$, two actions are allowed: elementary link generation and link swaps. 

\begin{figure*}[t] 
    \includegraphics[width=\textwidth]{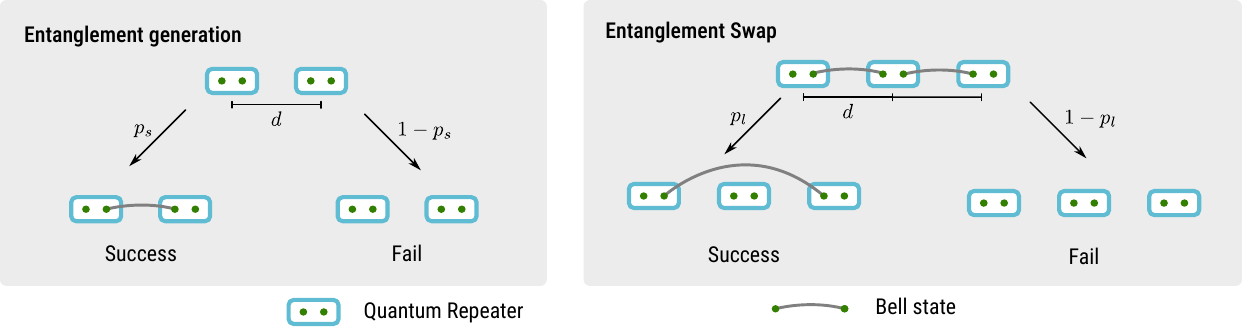}
    \caption{Left: Entanglement generation generates an elementary link between neighbouring nodes, separated by a distance $d$, with probability $p_e$ and fails with probability $1-p_e$. 
    Right: Entanglement swap merges two links into one with probability $p_s$ and fails with probability $1-p_s$. When it fails, all links involved in the swap are lost.}
    \label{fig:repeater actions}
\end{figure*}

Elementary link generation on a segment $i$, i.e. between nodes $i$ and $i+1$, probabilistically generates an elementary link $\{i,i+1\}$ between those nearest neighbours. 
These elementary links can then be probabilistically extended through link swaps~\cite{briegel_quantum_1998}. 
More precisely, when a node $i$ is linked with two other nodes, $j$ and $k$, an entanglement swap converts the two links $\{i,j\}$ and $\{i,k\}$ into a single link $\{j,k\}$, cf. Fig.~\ref{fig:repeater actions} for a schematic overview.
This process can then be iterated, enabling entanglement distribution over long distances, until end-to-end links are reached~\cite{briegel_quantum_1998}. As we only allow for the generation of elementary links, we will simply refer to elementary link generation as link generation. 

In our work, the link generation and swap actions can be performed regardless of the state of the quantum network. 
If a link generation action is applied to a segment, it first discards the current links to free up the corresponding qubits and then attempts the new link generation. 
For the swap action, if a node has less than two links, the end result is always a fully unlinked node. 

The linear homogeneous quantum repeater chains in this work are characterised by four parameters $(n, p_e, p_s, t_\text{cut})$, $n$ the number of nodes, $p_e$ the success probability for link generation, $p_s$ the success probabilities for the swaps and $t_\text{cut}$ the cutoff time.
The above parameters are motivated by physical processes.
As various experimental platforms only allow for probabilistic link generation and swapping.
We use the probabilities $p_s$ and $p_e$ to characterize them and assume for ease of exposition that they are the same for all nodes and independent of the underlying quantum states~\cite{azuma_quantum_2023}.
To each link, we associate an effective age that depends on how long the link has been present and how many times it has been swapped, to characterize its quality. 
The cutoff time $t_\text{cut}$~\cite{rozpedek_parameter_2018, li_efficient_2021, kamin_exact_2023, inesta_optimal_2023, haldar_reducing_2024} sets a maximum value on the allowed age of the links. 
A specific state or configuration of the quantum network is determined by the links that are present and the corresponding age. 
See Appendix~\ref{app:QN_states_and_actions} for more rigorous explanations of the above notions. The rest of the section is devoted to motivating the physical setting. 

The physical quantum state that is present after successful link generation is $\ket{\phi} := (\ket{00} + \ket{11})/\sqrt{2}$, a Bell pair state, which is maximally entangled. Links in this work are therefore also referred to as entanglement. 
Physically generating links on a segment $i$ can be achieved in various ways. 
We focus on the `meet-in-the-middle' scheme~\cite{duan_long-distance_2001, humphreys_deterministic_2018},
because of its low classical communication costs and entanglement generation being heralded, i.e.~the nodes are notified if entanglement has been generated.
In this scheme~\cite{duan_long-distance_2001, humphreys_deterministic_2018}, the two nodes create qubit-photon entanglement locally, followed by transmitting the photon to an interference station located precisely in between the nodes.
At the midpoint, the photons interfere and are subsequently detected, which entangles the photons if a specific detection pattern is observed.
The pattern is sent as a classical message to the nodes, confirming entanglement generation if successful.

The entanglement swap converts two Bell pairs into one by teleporting a qubit from one pair using the other Bell pair as resource~\cite{bennett_teleporting_1993}.
That is, the physical operation behind the entanglement swap takes two Bell pairs, one between qubits labelled by $A_2$ and $ B_1$ and one labelled by $B_2$ and $C_1$, and it measures out the two qubits labelled by $B_1, B_2$, so that the final result is a Bell pair between $A_2$ and $C_1$, modulo some local rotations that depend on the measurement outcome.
Afterwards, heralding messages, including the measurement outcomes, are broadcasted from node $B$, which contains qubits $B_1$ and $B_2$, to the other nodes to inform them whether qubit $A_2$ in node $A$ is linked to qubit $C_1$ in node $C$.
Finally, the measurement outcome corresponds to a local operation which $C$ performs to bring the entanglement back to a Bell pair, i.e. the entire operation maps
\begin{align}
    \frac{1}{\sqrt{2}}(\ket{00} + \ket{11})_{A_2B_1} \otimes \frac{1}{\sqrt{2}}(\ket{00} + \ket{11})_{B_2C_1} \nonumber \\ 
    \rightarrow \frac{1}{\sqrt{2}}(\ket{00} + \ket{11})_{A_2C_1}.
    \nonumber
    \label{entanglement_swap}
\end{align}

If the transmission speed (for both photons and classical messages) is denoted by $v$ and if $d$ is the internode distance, then for a homogeneous network the time taken for an elementary entanglement generation attempt is $d/v$ everywhere, the same as the communication time between nearest neighbours.
Regarding the entanglement swap, we use the common assumption that classical communication times are the dominant factor and therefore that the local swap operation at node $i$ is instantaneous~\cite{munro_inside_2015}.
For this reason and following convention~\cite{munro_inside_2015},
a single time step in this work is $d/v$, and all communication time is always an integer multiple of this unit. 

We assume that elementary-link generation yields a noisy Bell pair $\rho(p_{\textnormal{el}})$ for some $0 \leq p_{\textnormal{el}} \leq 1$, which is defined as
\[
\rho(a) := a \dyad{\phi} + (1-a)  \frac{\mathbb{1}}{4},
\]
where $\mathbb{1} / 4$ denotes the two-qubit maximally-mixed state.
Equivalently, $\rho(a)$ is a perfect Bell pair which underwent uniform depolarising noise with parameter $0\leq a \leq 1$~\cite{nielsen_quantum_2010}.
The state $\rho(a)$ has fidelity $\frac{1+3a}{4}$ with the perfect Bell pair $\ket{\phi}$, so $a$ is a direct indicator of the link's quality.
It can be shown that if a (noiseless) entanglement swap is performed on states $\rho(a)$ and $\rho(a')$, the resulting state is $\rho(a \cdot a')$~\cite{azuma_quantum_2023}.
Consequently, in the absence of other noise sources, the state of the link between the end nodes that an $n$-segment repeater chain produces is $\rho(p_{\textnormal{el}}^n)$~\cite{goodenough_noise_2024}.

When the quantum memories storing the qubits are imperfect, the quality of the links degrade over time.
We also model quantum-memory noise as a uniform depolarising channel, with parameter $a = e^{-t / T}$ with $t$ the time that the link has been stored in memory and $T \in (0, \infty)$ the memory coherence time, which indicates the quality of the memory.
One can derive that if a state $\rho(a)$ is stored in memory for time $t$, then the resulting state is $\rho(a  e^{-t / T})$, i.e.~the fidelity has decayed to $\frac{1 +3 a  e^{-t / T}}{4}$.
We refer to the duration $t$ that an elementary link is stored in memory as its `age', after which its state is $\rho(p_{\textnormal{el}}  e^{-t / T})$.
After entanglement swapping two links with age $t_{\textnormal{L}}$ and $t_{\textnormal{R}}$, respectively, the resulting state has parameter $\rho(p_{\textnormal{el}}^2  e^{-(t_{\textnormal{L}} + t_{\textnormal{R}}) / T})$ and thus we refer to $t_{\textnormal{L}} + t_{\textnormal{R}}$ as the `effective age' of the resulting link~\cite{reis_deep_2023}.

Adding quantum memory noise, the distributed state between the end nodes of the $n$-node quantum repeater chain will be 
\begin{equation}
\rho(p_{\textnormal{el}}^n \, e^{- t_{\textnormal{eff}} / T}),
\label{eq:end-to-end-state}
\end{equation}
where $t_{\textnormal{eff}}$ is the effective age of the end-to-end entanglement.

To mitigate the reduction of state quality due to memory noise, we impose a cut-off time $t_{\text{cut}}$~\cite{rozpedek_parameter_2018, li_efficient_2021, kamin_exact_2023, inesta_optimal_2023, haldar_reducing_2024}: when the effective age of the link surpasses $t_{\text{cut}}$, it is discarded, and the distribution of entanglement between the corresponding nodes is started anew. 
Thus enforcing a maximum value on $t \leq t_{\text{cut}}$ allows us to guarantee a maximum value on the effective age of the end-to-end link, and through Eq.~\eqref{eq:end-to-end-state}, a minimum value on the fidelity of the end-to-end state can be derived~\cite{inesta_optimal_2023}.
For this reason, as the quality of the link can be controlled through the $t_\text{cut}$ parameter, the $p_\text{el}$ parameter will not appear in our characterization of the repeater chain. 

Our goal in this work is to optimise quantum repeater protocols in the presence of classical communication time in terms  the delivery time, while guaranteeing some minimum quality on the end-to-end entanglement.
Since by setting a maximum for the cut-off time appropriately, a minimum fidelity is enforced, for the task of optimising the repeater protocol we thus only need to focus on minimising the average time at which end-to-end entanglement is delivered~\cite{inesta_optimal_2023}.

We finish by noting that the above noise model --- uniform depolarising noise --- can be generalised to arbitrary inhomogeneous Pauli noise, cf. Appendix~\ref{app:pauli_noise_model}. 
The main difference is that instead of only having one age $t$, one should keep track of three different weighted parameters.

\subsection{Related work} \label{previous works}

Significant research efforts have been vested towards the performance analysis of specific protocols for repeater chains and the optimization of specific parameters, see e.g.~\cite{azuma_quantum_2023} and references therein as well as more recent work~\cite{goodenough_noise_2024, coopmans_improved_2022, kamin_exact_2023, dai_entanglement_2021, guedes2024analysis}.

Recently, automated optimization over repeater policies has also been performed~\cite{shchukin_optimal_2022, inesta_optimal_2023,haldar_fast_2024, silva_optimizing_2021}.
Most of these works formalise the policy optimisation problem as a Markov decision problem~\cite{khatri_design_2022, shchukin_waiting_2019}.
Provably optimal delivery time policies have been found in~\cite{inesta_optimal_2023} using dynamic programming under certain assumptions, improved policies have been found using Q-learning in more general scenarios~\cite{haldar_fast_2024}, deep learning has been applied to optimise secret-key rates obtained using quantum repeaters~\cite{reis_deep_2023}, and genetic algorithms for optimising network parameters~\cite{silva_optimizing_2021}.
However, inclusion of classical communication effects remain elusive. 

To the best of our knowledge, quantum-repeater performance in the presence of classical communication delays was studied only very recently for the first time~\cite{haldar_reducing_2024}.
The authors analyse the performance of a family of protocols where the nodes hold multiple links with other nodes and decide which to swap based on given heuristics.
If there is at most a single link between any pair of nodes, as we consider in this work, the policies studied in \cite{haldar_reducing_2024} reduce back to the common swap-asap policy (see Sec.~\ref{fixed_policies}).
In this work, we perform automated optimisation, using reinforcement learning, over quantum repeater policies with classical communication delays.
Instead of requiring the policy to wait for classical communication to finish, we allow it to take actions sooner.
This greedier approach is actually beneficial in parameter regimes where the swap and entanglement generation actions are likely to succeed. 

\section{Delays through classical communication effects}\label{CC_delays}

In order to define our Markov decision process formulation for the policy optimisation problem, we need to specify our model for the classical communication effects.

Classical communication effects have two main effects.
Firstly, they cause delays in performing actions.
If we consider a single agent that controls the entire network, then actions at more distant nodes are executed with  increased delays.
Secondly, classical communication effects give rise to delays with which the results, and thus information about the state of the network, can be retrieved.
Intuitively, an agent experiencing classical communication effects must coordinate their actions such that the different delays of various actions are taken into account. 
Additionally, they must choose which results to wait for and which results not to wait for, with the benefit of faster response times. 

\begin{figure}[t]
\centering\includegraphics[width=\columnwidth]{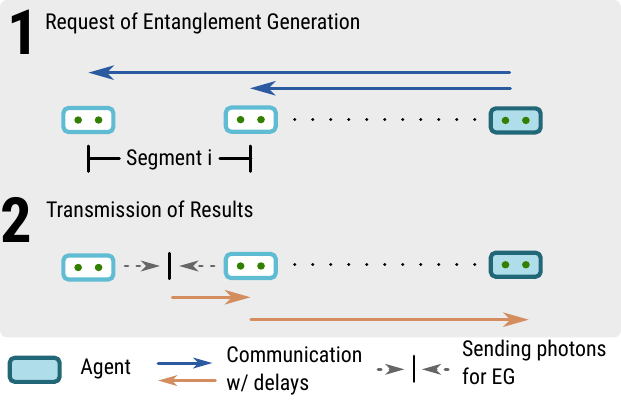}
    \caption{For a segment where the leftmost node is a distance $d$ away from the agent, sending the message takes $d$ time steps. 
    Then sending two qubits to the middle takes $d/2$ time steps and sending the result from the middle of the segment to the agent takes another $d-\frac{1}{2}$ time steps. 
    In total, sending out the actions and getting back the result thus takes $2 d$ time steps.}
    \label{EG_CC}
\end{figure}

The delay in performing entanglement generation over segment $i$ selected by an agent located at node $k$ is equal to 
\begin{equation}
    \Delta_{\text{EG}}(i,k) := \text{max}(|k-i|,|k-i+1|). \label{EG delay}
\end{equation}

We take the most distant node, because in link generation, both nodes of the segment have to emit a photon simultaneously, and for that, both nodes must have received the instruction to do so, cf. panel $1$ in Fig.~\ref{EG_CC}. 
The time it takes for the agent to receive the result also corresponds to $\Delta_{\text{EG}}(i,k)$.
This is because once the nodes have emitted the photons, it takes half a time step for them to reach the midpoint, and then another half a time step for the heralding message to be sent to the closest node. 
From the closest node, the heralding message travels the rest of the distance to the agent, which in total is equal to the distance between the agent and the most distant node, cf. panel $2$ in Fig.~\ref{EG_CC}. 
We write $\Delta_{\text{EG}}(k) := \text{max}_i\left( \Delta_{\text{EG}}(i,k) \right)$ to denote the entanglement generation delay to the farthest away node. 

\begin{figure}[t]
    \centering
    \includegraphics[width=\columnwidth]{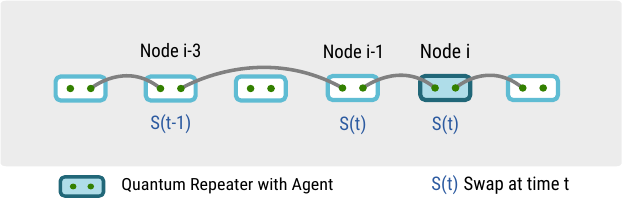}
    \caption{A swap is performed at time step $t$ at node $i$ where the agent is located. 
    The agent has also sent out swap instructions in the previous time steps such that node $i-1$ also performs a swap at time step $t$ and node $i-3$ performed a swap at time $t-1$. 
    To know whether its swap at node $i$ has succeeded, it also needs to know whether the swap at node $i-1$ and node $i-3$ has succeeded or not. If the swap at node $i-3$ failed, for example, then the swap at node $i-1$ will fail as well. 
    This in turn would mean that the swap at node $i$ also fails automatically.
    }
    \label{Swap_CC}
\end{figure}

For the entanglement swap, the delay in executing the action at node $i$ sent by the agent at node $k$ is
\begin{equation}
    \Delta_{\text{swap}}(i,k) :=|k-i|. \label{swap action delay}
\end{equation}
To know whether a swap has been successful, however, the measurement outcome alone is not sufficient. 
Additional information about whether the swapped node actually had two links at the moment of the swap is also needed. 
If fewer than two links are present, the swap will always fail. 
It is not always obvious if a node had two links the moment it was swapped because multiple swaps can be performed on different nodes simultaneously in each time step, cf. Fig.~\ref{Swap_CC} for an example. 
If the nodes on which the swaps are performed are all linked together, then a failure in one of the swaps would lead to all of the links being discarded. 
Due to classical communication delays, the nodes have to wait for the swap measurement outcomes of all relevant nodes to verify if their swap has succeeded or not. 
Instead of keeping track of which results are relevant for which swaps, we wait for information to propagate from the most distant non-end node which takes
\begin{equation}
    \Delta_{\text{swap-result}}(k) :=\text{max}(k-1,n-2-k) \label{swap result delay}
\end{equation}
time steps. 
In that case, the agent has had enough time to gather results from every non-end node and thus waited long enough to know whether the swap has been successful or not. 
For Eq. \ref{EG delay}, \ref{swap action delay} and \ref{swap result delay}, we omit the position of the agent $k$ for brevity when it is clear from the context. 

\section{Markov decision process formulation}\label{MDP formulation}

The Markov decision problem formalism is a powerful way to formulate policy optimization problems, for which many methods have been developed, such as dynamic programming methods, temporal-difference learning and deep reinforcement learning, cf. Ref.~\cite{sutton_reinforcement_2020} for more details. 
Formally, a Markov decision process (MDP) is a $4$-tuple $(\mathcal{A},\mathcal{S},P,R)$. 
Here, $\mathcal{A}$ is the set of all possible actions $a \in \mathcal{A}$ that the agent is allowed to take, $\mathcal{S}$ is the set of all possible states or observations $s \in \mathcal{S}$ that the agent can observe, 
$P(s',r|s,a)$ is the transition probability of the environment and $R$ determines the values for the reward function $r(s,a)\in R$.
In an MDP, in each step, the agent takes an action $a$ based on its current observation $s$. 
After the action, the environment, transitions to a new state $s'$ and gives a reward $r(s,a)$ back to the agent according to the transition probability $P(s',r|s,a)$. 
The goal is for the agent to find the policy, i.e.~probability distribution $\pi(a|s)$, such that the cumulative rewards are maximised. 

\begin{figure*}[t]
\includegraphics[width=\textwidth]{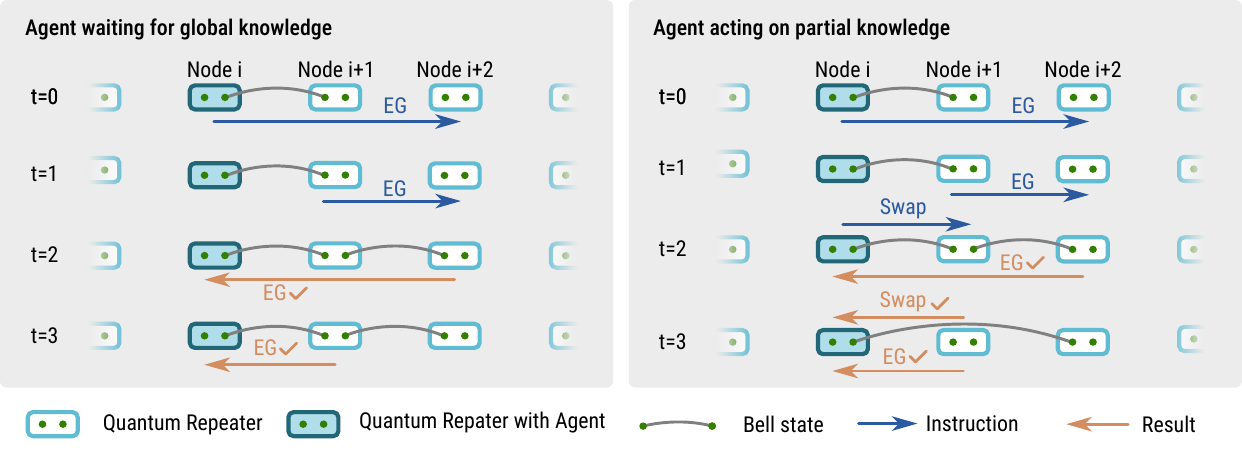}
\caption{It takes one time-step for information to move
from one node to the nearest neighbour. 
At $t=0$, node $i$ and $i+1$ are already entangled and the agent decides to attempt entanglement generation between nodes $i+1$ and $i+2$. 
An agent that does not wait for the result of the entanglement generation action can choose to send a swap instruction to node $i+1$ at $t=2$. 
If both the swap and the generation actions are successful, the agent that does not wait for global information will be able to establish a longer link faster than the agent that waits for global information.}
\label{fig:cc protocols}
\end{figure*}
 
In our MDP, within one time step, we allow each segment to attempt entanglement generation once and each non-end node to attempt a swap once.  
After each time step, the age of all links are increased by one.
More specifically, we split each time step into two rounds, one dedicated to each of the two types of actions.
This has the added benefit of reducing the size of our action space.
Instead of choosing all of the segments on which link generation will be attempted and all of the nodes where swaps will be performed at once, we will reserve every even round for swap actions and every odd round for link generation actions.
Though swaps can in principle be performed multiple times within a time step, this is not restrictive as doing multiple swaps on the same node, without entanglement generation in between, does not result in more possible transformations. 
Conversely, multiple attempts of entanglement generation on the same segments are not possible within one time step. 
Within a single time step, it is sufficient to allow for one round of entanglement generation actions followed by one round of swap actions.

We explicitly allow actions to be sent out each time step, without waiting for the results. 
This has the benefit that if the node is likely to be linked, it can be extended without waiting for the verification, 
cf. Fig~\ref{fig:cc protocols} for an example. 

If there were no classical communication effects, one can in principle perfectly reconstruct the current quantum network state from past actions and results. 
When classical communication effects are present, the delay in the results means that, in general, at the current time step there is not enough information to reconstruct such a representation of the current state. 
In this work, instead of attempting to make an approximate reconstruction given incomplete information, we take the history up to a chosen cutoff as the observation directly. 
We thus construct our MDP informally as follows. 
For a more detailed specification, we refer to Appendix~\ref{app:mdp}.

\paragraph{Action space:} Depending on the parity of the round, the agent can either select segments or nodes to which entanglement generation or swap instructions are sent respectively.
There are $n-1$ segments on which entanglement generation can be attempted and $n-2$ non-end nodes on which swaps can be attempted.
To this extent, we construct our action space to have size $2^{n-1}$ such that there is an element $a \in \mathcal{A}$ for each possible combination of segments on which entanglement generation can be performed. 
For the swap actions, as there are $n-2$ non-end nodes on which swaps can be attempted, this thus means that this choice results in a redundancy in the swap actions. 
Our scaling is comparable to Refs. \cite{inesta_optimal_2023, reis_deep_2023, haldar_fast_2023}, where the actions are similarly constructed by assigning booleans to nodes or segments. 

\paragraph{Observation space:}
The observation that the agent receives in each round is a history of all the actions and the corresponding results of the past $t_\text{cut}$ time steps.
We only keep track of the past $t_{\text{cut}}$ time steps, because any action or result that happened before that corresponds to a link that has already been discarded. 

\paragraph{Reward function:}
The figure of merit that we optimise for is the end-to-end delivery time, as with an appropriately chosen cutoff time, the desired end-to-end fidelity can guaranteed. 
The rewards are chosen such that the delivery time is minimised when the reward is maximised.
However, in general, the agent is not able to observe that end-to-end entanglement has been reached the moment it occurs.
This is due to the delay in the results.
In order for an episode to terminate, we require that end-to-end entanglement must be held for $2 \text{max}(k, n-1-k)$ rounds, such that the agent has had enough time to verify that the state is indeed end-to-end entangled. 
When the terminal state is reached, the reward function evaluates to $r = 0$. 
In all other rounds, the reward function evaluates to $r = -1$.
This enforces that the cumulative reward is minimised when the delivery time is maximised. 
  
\paragraph{Environment:} 
As entanglement generations and swaps only succeed probabilistically, the dynamics of the environment are stochastic.
The probability to go from one history to the next one is determined directly by the success probabilities corresponding to results that are received in the current round. 

Our formulation is most similar to the ones found in Refs.~\cite{inesta_optimal_2023, haldar_fast_2024}. 
The main difference is that we are using a history as the observation rather than a quantum network state. 
Without classical communication effects, we could instead also use the current state of the quantum network as the observation, in which case our MDP would be analogous to the aforementioned ones. 
In Ref.~\cite{inesta_optimal_2023}, entanglement generation is always attempted. 
In our work, the policy can decide to wait and attempt entanglement generation at a later point in time. 

\section{Fixed policies}\label{fixed_policies}

When classical communication effects are absent, the swap-asap policy~\cite{coopmans_tools_2021, kamin_exact_2023, goodenough_noise_2024} is commonly studied as it is optimal in the case of $p_e= p_s=1$, and also when only swaps are deterministic ($p_s = 1$) but entanglement is never discarded through a cut-off~\cite{kamin_exact_2023, goodenough_noise_2024}. 
For scenarios where the swap-asap is not optimal, it serves as a baseline against which other policies are compared~\cite{inesta_optimal_2023, haldar_fast_2024}.
Additionally, it also forms a useful starting point for designing more complex multiplexed protocols~\cite{haldar_reducing_2024}. 
In particular, in the scenario where each node only has two qubits, the improved policies found in~\cite{haldar_reducing_2024} reduce to the swap-asap policy. 

The standard swap-asap policy where classical communication effects are not taken into account will be referred to as the instantaneous swap-asap policy in our work. 
In this policy, there are no delays with which the swap-asap actions are performed nor with which the state of the quantum network is known.
It is thus expected that the delivery time of such an instantaneous policy is lower than even the best policy with classical communication effects. 
In addition to providing a lower bound, comparing the difference in delivery time between them allows us to highlight the importance of explicitly considering classical communication effects. 

\begin{figure}[t]
    \centering
    \includegraphics[width=\columnwidth]{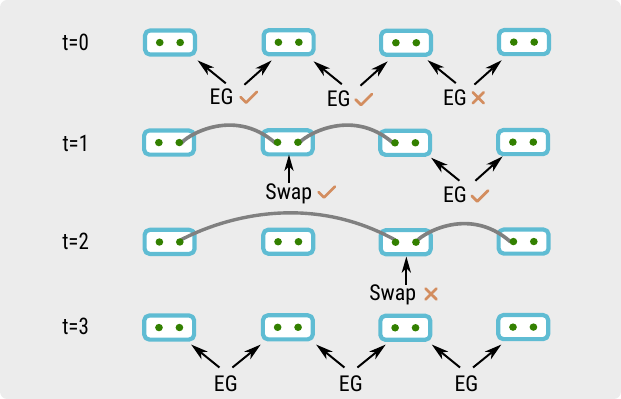}
    \caption{Example of the swap-asap policy with instantaneous communication on a 4-node chain. At $t=0$ there are no links, and all segments attempt link generation, of which the zeroth and first segment succeed. At $t=1$, since node $1$ has two links, it attempts a swap which succeeds. Segment $2$ attempts link generation again, which now also succeeds. At $t=2$, node $2$ attempts a swap which fails. All links are discarded and link generation is attempted again at each segment.}
    \label{fig:swap_asap_example}
\end{figure}

\paragraph{Instantaneous swap-asap policy:} 
In this policy, in each time step, the policy first attempts swaps on all non-end nodes that have two links. 
After that, entanglement generation is attempted on all segments where both qubits are not linked yet. 
For a schematic example of the instantaneous swap-asap policy, cf.~Fig.~\ref{fig:swap_asap_example}.
After the entanglement generation action, the time is increased by one. 

In this work, we make multiple modifications to the swap-asap protocol to take into account classical communication effects in various ways. 
In addition to being easier to analyse than reinforcement learning policies, these will serve as benchmarks against which we can compare our optimised MDP policies. For a more detailed description of policies described in this section, cf. Appendix~\ref{app:swap_asap}.

\paragraph{Wait-for-broadcast (WB) swap-asap policy:} 
Here we provide a direct generalization of the swap-asap policy where classical communication effects are taken into account. 
This is a variant of the swap-asap policy where the actions are selected by a single agent located at node $k$.
In each action round, it performs either only entanglement generation or swap actions. 
If the previous actions were swaps, it performs entanglement generation actions in the current round and vice versa. 
After a swap round, the policy waits $2  \Delta_{\text{swap-result}}(k)$ time steps and after an entanglement generation round it waits $2  \Delta_{\text{EG}}(k)$ time steps. 
This is to ensure there is enough time for the policy to send the actions to any combination of nodes or segments and to retrieve the corresponding results so that the agent has full information of the current state.
Depending on the round, the agent attempts entanglement generation on all free segments or swaps on all non-end nodes with two links. 
This policy is included because a reinforcement learning agent that has access to the full history should be able to perform at least comparably.  
In the worst case, the reinforcement learning agent can choose to wait for full global information and select actions according to this WB swap-asap policy.

We note in particular that the WB swap-asap policy is no longer optimal at $p_s, p_e =1$ when classical communication effects are present. 
The WB swap-asap policy needs $2 \Delta_{\text{EG}}(k) + 2 
\Delta_{\text{swap-result}}(k)$ time steps to deliver end-to-end links. 
It first sends out entanglement generation actions to all segments and once it has verified that all segments are linked, it sends out swap actions and waits for the corresponding results.  
When $p_s, p_e = 1$, the optimal policy with a global agent located at node $k$ is able to generate end-to-end links in $\Delta_{\text{EG}}(k)+1+\Delta_{\text{swap-result}}(k)$ time steps, as it needs $\Delta_{\text{EG}}(k)$ time steps for sending entanglement generation actions to all nodes, then it needs 1 time step to perform the entanglement generation actions, after which the swap actions are directly performed. 
Then it needs another $\Delta_{\text{swap-result}}(k)$ time steps to collect all of the results. 
As actions in this scenario succeed with unit probability, collecting the results is not strictly needed, but we still include them in our arguments for consistency with policies at lower success probabilities where the above argumentation does not hold. 
By a continuity argument, for sufficiently high success probabilities, policies yielding faster expected delivery times than the WB swap-asap are expected to be found. 

More generally, we expect to find even better policies if we allow for local policies.
By this we mean policies where each node selects its swap action itself and its entanglement generation action together with the corresponding neighbour. 
The advantages of this policies are built on the idea that neighbouring information might be more valuable than distant information. 
Additionally, the delay in performing the actions are minimal. 
Based on these ideas, we construct the predictive swap-asap policy. 

\paragraph{Predictive swap-asap policy:} In this swap-asap adaptation, the swap and entanglement generation actions are directly selected by the nodes executing them.
Similarly to the MDP formulations, the time steps are split into two rounds, one for each type of action.  
After each action, the policy updates its view of the state probabilistically depending on $p_e$ and $p_s$. 
For example, if node $3$ decides to perform a swap, it also updates its view of the state as if the swap at node $3$ has succeeded $p_s$ of the times and as if it has failed $1-p_s$ of the times. 
This policy is included because in the scenario where swap and entanglement generation both succeed with unit probability, the prediction is always correct, and the performance thus matches the one of the instantaneous swap-asap protocol, up to end-to-end verification time.

\section{Reinforcement learning algorithm}\label{RL_algorithm}

For the policy optimization task in the MDP formulation, we employ reinforcement learning methods.
We choose this approach due to the exponential size of the observation space that we optimise over, which is upper bounded by $6^{t_\text{cut}(n-1+n-2)}$.
The $6$ corresponds to the $2\cdot 3$ options of sending or not sending an action, combined with having a positive result, negative result or no result. 
The terms in the exponent represent 
the $n-1$ segments on which entanglement generation can be attempted and $n-2$ non-end nodes on which swaps can be attempted, multiplied the past $t_\text{cut}$  time steps that we keep track of.  
Due to this size, a dynamic programming or tabular approach to optimise policies does not seem suitable and instead we opt for a deep reinforcement learning method.
The specific algorithm we use is the Proximal Policy Optimisation (PPO) algorithm~\cite{raffin_stable-baselines3_2021}, a policy gradient method,
which in practice has been shown to be particularly robust~\cite{sivak_model-free_2022, yu_surprising_2022, zhu_quantum_2023, reis_deep_2023, zen_quantum_2024, nagele_tackling_2024, nagele_optimizing_2024}

\section{Numerical experiments} \label{NumRes}

\begin{figure*}[t]
\includegraphics[width=\textwidth]{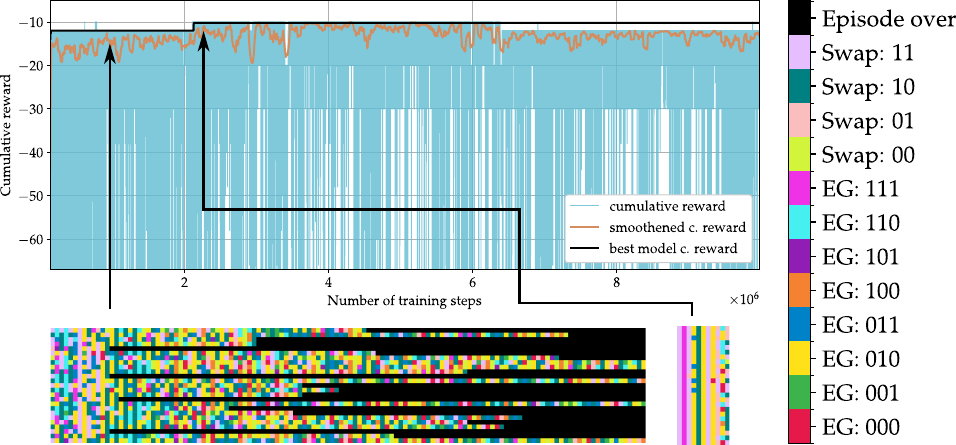}
\caption{
\emph{Top:} Training curve of the reinforcement learning agent for $(n, p_e, p_s, t_\text{cut}) = (4, 1, 1, 12)$. 
The black curve is obtained by periodically saving the best-trained model so far and plotting the associated reward.
\emph{Bottom:} 
The two panels show the actions according to the policies at different stages during training. 
Each row in the heatmap corresponds to an independent episode.
The time axis runs from left to right and each pixel signifies one round in the Markov decision process.
The colour of each pixel corresponds to a specific action taken in a given round (see legend).
In the legend, we use binary strings to represent the entanglement generation and swap actions, e.g.~the label \enquote{EG:$101$} corresponds to entanglement generation actions on elementary links $0$ and $2$.
Similarly, the action \enquote{Swap:$11$} signifies a swap on nodes $1$ and $2$.
Note, that a swap on the first and the last node is impossible, thus the condensed notation for the swap.
The policy corresponding to the left panel is obtained after 91590 training steps and the other policy after 2131191 training steps. The episodes of the leftmost heat map have been trimmed to fit the figure. }
\label{fig:policy_interpretation}
\end{figure*}

To numerically estimate the delivery times, we perform Monte Carlo simulations of each policy and average over the different episodes. 
We focus on repeater chains of $4$ nodes where the agent is located at node $2$.
This is already sufficient for illustrating the relevance of classical communication effects. 
Additionally, we expect the advantages found for smaller networks to be extendable to larger networks through nested policies, as proposed in \cite{haldar_fast_2024}.
In these nested policies, the network is divided into partitions, and end-to-end links are first established within the separate partitions. 
Then the partitions are connected together to establish a link over the full length of the network. 
As faster end-to-end links can be delivered within partitions of 4 nodes, faster end-to-end links spanning the entire network can be delivered as well. 
The swapping probabilities $p_s$ that we considered are $0.5$, $0.75$ and $1$, motivated by linear optics setups~\cite{grice2011arbitrarily}. For $p_e$ we considered probabilities between $0.1$ and $1$ in steps of $0.1$. 
The cutoff time is set at $t_\text{cut}=12$.

For the instantaneous swap-asap, as the delivery time is low, we fully simulate the policy for all parameters.  
To estimate the delivery times of the WB swap-asap on the other hand, we simulated the instantaneous swap-asap, and multiplied the delivery time by $\Delta_{\text{EG}}(2) + \Delta_{\text{swap}}(2) + \Delta_{\text{swap-result}}(2) = 6$, to account for the broadcast delay.
To account for this multiplication, the cutoff time during the simulation is set at $2$ so that after multiplication it matches the cutoff time of the other policies. 
For the predictive swap-asap and the reinforcement learning, full simulations were also performed.
Due to the high simulation times at low success probabilities, however, data is omitted if the simulation took more than $2$ seconds or $5\cdot 10^4$ steps per episode to simulate on average.
To obtain the reinforcement learning policy, for each combination of $(p_s, p_e)$, we trained 20 different agents and selected the best one.  

For the point $p_s=p_e = 1$, Fig.~\ref{fig:policy_interpretation} shows that the optimal policy is much more structured than the suboptimal policy during training . 
The agent learns to prioritise and discard actions, leading to a simplified policy.
This opens up the possibility of interpreting the resulting policy.
While the left heat map in Fig.~\ref{fig:policy_interpretation} appears mostly unstructured, the right one has strong vertical lines across almost all rounds.
This means that in each step the same actions are performed, independent of the episodes. 
In fact, the strategy found in the right heat map is the optimal strategy as it takes 11 rounds, which is 5 time steps. 
We note that this optimal strategy is not unique however. 
The action \enquote{Swap:$11$} is for example performed in the zeroth, first and third time step in almost all episodes, though in principle only one swap per non-end node is needed. 
In fact the swap actions selected in the first time step even discards the link at a segment $1$.
However, the link at segment $1$ is restored again in the same time step as the link at segment $0$ is created. 
Therefore, the end-to-end delivery time is not affected.
More generally, the optimal policy allows for some freedom in the choice of actions that are selected before the $\Delta_\text{EG}$'s time step, as long as enough links to span the length of the network are present at time $\Delta_\text{EG}$.  
Similarly, all actions that do not affect the end nodes when an end-to-end link has been created are allowed as well. 
This also explains the variation in actions in the last two rounds, as destructive actions would not arrive in time due to the classical communication delays. 

For lower success probabilities, we observe a similar pattern for the obtained policy after training, cf. App.~\ref{app:heatmaps}.
Across different episodes, the same actions are frequently taken in the same time step. 

\begin{figure*}[t] 
\includegraphics[width=\textwidth]{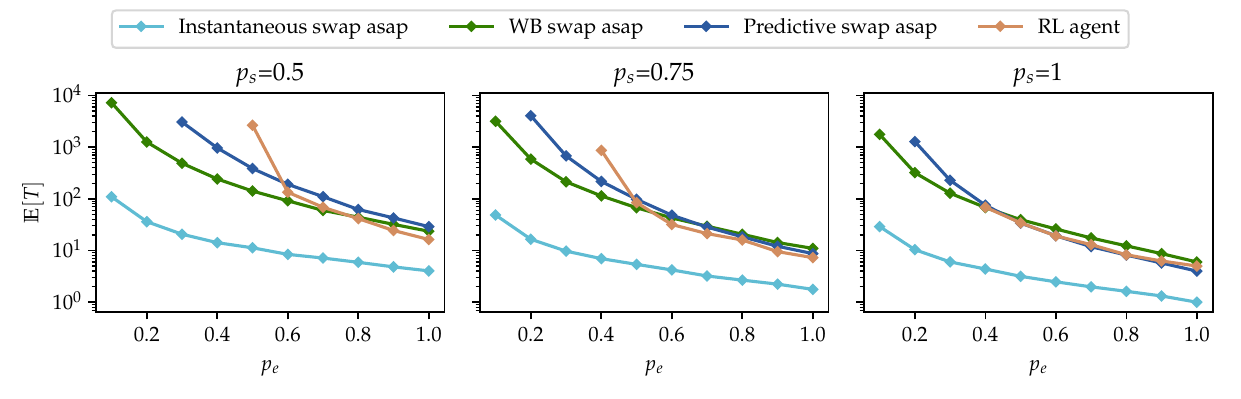}
\caption{
Plots of the expected delivery time of various protocols at $n=4$ and $t_\text{cut} = 12$ for various values of $p_s$ and $p_e$. 
For the WB swap-asap and the RL setting, which use global agents, we have selected to put the agents on node $2$. 
It shows that at high probabilities, the RL and the predictive swap-asap policy are able to outperform the WB swap-asap. 
Points where the RL and predictive swap-asap policy took more than $2$ seconds per episode to simulate or where the expected delivery time was larger or equal to $5\cdot10^4$ time steps were omitted.}
\label{fig:results}
\end{figure*}

Our main results can be found in Fig.~\ref{fig:results}.  
 Policies with classical communications effects have higher delivery times than the instantaneous swap-asap (light blue curve), by roughly one order of magnitude or more.
 This demonstrates the importances of considering classical communication effects for realistic scenarios.
Additionally the predictive swap-asap (dark blue curve) and the reinforcement learning policy (orange curve) have lower delivery times than the WB swap-asap (green curve) in the high parameter regime.

We observe from Fig.~\ref{fig:results} that at the point where $p_s= p_e = 1$, the predictive swap-asap is able to achieve the lowest optimal delivery time as allowed within classical communication constraints. 
The predictive swap-asap policy delivers end-to-end entanglement in $4$ time steps, one time step for performing entanglement generation on each segment, after which swaps are performed on all non-end nodes. 
Then there are $3$ time steps for the end-to-end communication to take place so that all nodes can verify whether end-to-end links have indeed been established or not. 
Similarly, as noted before in Fig. \ref{fig:policy_interpretation}, the reinforcement learning agent is able to achieve the optimal global policy. 

Fig.~\ref{fig:results} also shows that the advantage that the predictive swap-asap has over the WB swap-asap decreases with decreasing success probabilities. 
This is expected as only at the point where all actions succeed, the predicted state of the policy is the same as the real state. 
In the quadrant $0.5 \leq p_s \leq 1$ and $0.5 \leq p_w \leq 1$, as the success probabilities decrease, the probability that the predicted result and the real result are equal to each other also decreases.
This results in increasingly incorrect predictions, misleading the nodes into selecting potentially harmful actions, e.g. swapping a node that only has one link.
At even lower probabilities of $p_s$ and $p_e$, we attribute the further increase in delivery time mainly to increasingly likely failures of the actions. 

For the reinforcement learning policy, the trend is comparable to that of the predictive swap-asap. 
However, as the reinforcement learning policy is expected to be able to learn a policy similar to the WB swap asap policy, we attribute decreasing performance mainly to trainability challenges.

\section{Conclusion and outlook} \label{conclusion outlook}

In this work, we have investigated the effects of classical communication delays on entanglement delivery policies. 
We first proposed a predictive local policy where the delays in performing the actions take at most $1$ time step.
We then used this as a benchmark to compare our global reinforcement learning agent against, which experiences delays that scale with the distance between itself and the node the actions are sent to. 
We focused on homogeneous equidistant repeater chains and we show that, already for $4$ nodes, advantages of our methods are found in the high success probability regime.
The advantages of this policy can be extended further to larger networks using a nested scheme, as was proposed in Ref. \cite{haldar_fast_2024}.
The decreasing performance at lower success probabilities can at least be partly attributed to trainability challenges of the reinforcement learning agent.

For the implementation in our current work, we have focused on the simpler representation of the observation, which is a history of actions and the corresponding results. 
For future work, it would be interesting to see if storing a reconstructed state of the quantum network instead of the history of actions could result in improved training performances, as it reduces the size of the observation space.

Another direction for future work is to further explore local policies.
As demonstrated by the predictive swap-asap, local policies offer advantages over global policies when actions succeed deterministically.  
Optimal local policies are expected to outperform optimal global policies in non-deterministic parameter regimes as well for two reasons.  
First, the time it takes to perform each of the actions at each node is constant for local policies, rather than scaling with the distance between the agent and said nodes for global policies. 
Additionally, information about neighbouring nodes, that are not at the location of the global agent, can also be gathered more quickly due to the smaller distances between them.
Using multi-agent reinforcement learning~\cite{busoniu_multi-agent_2010, zhang2021multi, yu_surprising_2022} to optimize over local policies, greater advantages are expected to be found.

To conclude, we have shown that by allowing policies to act with partial information when classical communication effects are taken into account, faster policies can be found. 
The proposed policies outperform a direct generalisation of the well-studied swap-asap policy, the wait-for-broadcast swap-asap policy.
Additionally, the policies found through reinforcement learning prove to be interpretable. 
As analytic investigations become increasingly challenging for more realistic scenarios, reinforcement learning methods provide a suitable alternative. 
Our work provides a first glance at the advantages that can be gained through using partial information policies and motivates further research in this direction.

\section*{Code availability}

The code used in this project can be found at: \url{https://github.com/janli11/QuantumNetworkGlobalPolicyWithCC}

\acknowledgements
T.C. acknowledges the support received through the
NWO Quantum Technology program (project number NGF.1582.22.035). 
J.L., P.E., J.T. and EvN acknowledge the support received by the Dutch National Growth Fund
(NGF), as part of the Quantum Delta NL programme. 
P.E. acknowledges the support received through the NWO-Quantum Technology
program (Grant No.~NGF.1623.23.006).
J.T. acknowledge the support received from the European Union's Horizon Europe research and innovation programme through the ERC StG FINE-TEA-SQUAD (Grant No.~101040729). 
This publication is part of the `Quantum Inspire - the Dutch Quantum Computer in the Cloud' project (with project number [NWA.1292.19.194]) of the NWA research program `Research on Routes by Consortia (ORC)', which is funded by the Netherlands Organisation for Scientific Research (NWO).

The views and opinions expressed here are solely
those of the authors and do not necessarily reflect those of the funding institutions. Neither
of the funding institutions can be held responsible for them.

\bibliography{references}

\appendix

\section{Quantum network states and actions}\label{app:QN_states_and_actions}

What follows in this section serves as a complementary description of the quantum network state and the entanglement generation and swap actions. 
It gives a more rigorous description to what is described in the main text.

\subsection{The state of a quantum network}

The state $\sigma \in \Sigma$ of a quantum network is characterised by a triple $(n,\Lambda,\tau)$. 
Here $n \in \mathbb{N}$ is the number of nodes in the linear network, where each node is labelled by an integer $i\in [n]$. 
The set $\Lambda$ is the set that contains all current links $\lambda_{(i,j)} = \{i,j\}$ between nodes $i$ and $j$ of the quantum network, and $\tau: \lambda_{(i,j)}\rightarrow \mathbb{R}$ is the age of the link. 
The state may thus equivalently be thought of as a weighted undirected graph, where $n$ denotes the nodes, $\Lambda$ the edges and $\tau$ the weights. 

\subsection{Actions}
\textbf{Link generation:} a link generation attempt between nodes $i$ and $i+1$ is denoted as $e_i$. 
When the link generation attempt is applied, it first removes all links connected to qubits at segment $i$, i.e. it removes all the links $\lambda_{i,j}$ with $j>i$ and $\lambda_{i+1,k}$ with $k<i+1$. 
Then it adds link $\lambda_{i,i+1}$ to $\Lambda$ with probability $p_e$ and nothing to $\Lambda$  with probability $1-p_e$. 
The age of the generated link is set to $0$.

\textbf{Swap action:} We denote a swap on node $i$ as $s_i$. It is a map that takes the quantum network from one state to another, i.e.~$s_i: \sigma \in \Sigma \rightarrow \sigma' \in \Sigma$. For a state $\sigma$ where node $i$ has two links, i.e.~it contains $\lambda_{(i,j)}$ and $\lambda_{(i,k)}$ where $j \neq k$, it maps $\sigma = (n,\Lambda=\{...,\lambda_{(u,v)},\lambda_{(i,j)},\lambda_{(i,k)},\lambda_{(x,y)},...\},\tau)$ to $\sigma' = (n,\Lambda'=\{...,\lambda_{(u,v)},\lambda_{(j,k)},\lambda_{(x,y)},...\},\tau')$ with probability $p_s$ and to $\sigma' = (n,\Lambda'=\{...,\lambda_{(u,v)},\lambda_{(x,y)},...\},\tau')$ with probability $1-p_s$.
It either merges $\lambda_{(i,j)}$ and $\lambda_{(i,k)}$ into a single link $\lambda_{(j,k)}$ with probability $p_s$ or it removes $\lambda_{(i,j)}$ and $\lambda_{(i,k)}$ from the set $\Lambda$ with probability $1-p_s$.
If the swap is successful, the age of the new link is the sum of the two consumed links $\tau'(\lambda_{j,k}) = \tau(\lambda_{i,j}) + \tau(\lambda_{i,k})$.
When node $i$ does not have links $\lambda_{(i,j)},\lambda_{(i,k)}$ where $j \neq k$, all links, if any, with the index $i$ are removed from $\Lambda$.

\section{Pauli noise model}\label{app:pauli_noise_model}

For simplicity we assumed in the main text that the only source of noise was depolarising noise. We will show here, in a similar fashion as was done in~\cite{goodenough_noise_2024}, that one can extend our analysis to arbitrary inhomogeneous Pauli noise, i.e.~noise maps of the form $\mathcal{N}(\rho) = \sum_{P\in \lbrace{ I, X, Y, Z\rbrace}}p_P P\rho P^\dagger$. In particular, we show that one can define analogous parameters that capture the quality of the underlying state.

First off, using the transpose trick~\cite{wilde_classical_2019} it is always possible to move Pauli noise from one side of a maximally entangled state to the other side, such that we can restrict ourselves to Choi states of Pauli channels. Second, swapping two Choi states of Pauli channels $\mathcal{N}_1, \mathcal{N}_2$ (with associated probabilities $p_{1, P}$ and $p_{2, P}$) yields a Choi state of the composition of the channels $\mathcal{N}_1\circ \mathcal{N}_2$, independent of the Bell state measurement outcome (see for example~\cite{haldar_fast_2024, haldar_reducing_2024, goodenough_noise_2024}). We will thus focus in the remainder of this section on the composition of Pauli channels.

The composition of two Pauli channels is again a Pauli channel, and can be naively calculated by summing the probabilities that would lead to applying a certain Pauli operator $P$. As an example, the probability of applying $X$ for the composition of $\mathcal{N}_1$ and $\mathcal{N}_2$ is given by $p_{1,I} p_{2,X}+p_{1,X} p_{2,I}+p_{1,Y} p_{2,Z}+p_{1,Z} p_{2,Y}$. Such sums quickly become unwieldy when extending to multiple swaps, and it is not clear from such sums how the quality decays as noise accumulates.

However, one can interpret such sums as a type of convolution over the Pauli group without phases~\cite{shahbeigi_log-convex_2021, goodenough_noise_2024}. By then applying a Fourier transform, the convolution turns into point-wise multiplication, and one recovers the exponential decay one expects. More explicitly, the Fourier transform is given by the following linear invertible map,

\begin{align}
\lambda_1 =~& p_I + p_X + p_Y + p_Z = 1\ ,\\
\lambda_2 =~& p_I + p_X - p_Y - p_Z\ ,\\
\lambda_3 =~& p_I - p_X + p_Y - p_Z\ ,\\
\lambda_4 =~& p_I - p_X - p_Y + p_Z\ .
\end{align}

Thus, to calculate the resultant Pauli channel after composing Pauli channels $\mathcal{N}_1, \mathcal{N}_2, \ldots, \mathcal{N}_N$, one first calculates the above four parameters $\lambda_{i, 1}, \lambda_{i, 2}, \lambda_{i, 3}, \lambda_{i, 4}$ for each channel $\mathcal{N}_i$. Secondly, one calculates the four products $\prod_{i=1}^ N\lambda_{i, j} \equiv \overline{\lambda}_j$, for $1 \leq j\leq 4$, which are exactly the $\lambda$ parameters of the final Pauli channel (since the Fourier transform transforms convolution into point-wise multiplication). Finally, inverting the above Fourier transform yields

\begin{align}
\overline{p}_I =&~ \frac{\overline{\lambda}_1+\overline{\lambda}_2+\overline{\lambda}_3+\overline{\lambda}_4}{4}\ ,\\
\overline{p}_X =&~\frac{\overline{\lambda}_1+\overline{\lambda}_2-\overline{\lambda}_3-\overline{\lambda}_4}{4}\ ,\\
\overline{p}_Y =&~ \frac{\overline{\lambda}_1-\overline{\lambda}_2+\overline{\lambda}_3-\overline{\lambda}_4}{4}\ ,\\
\overline{p}_Z =&~ \frac{\overline{\lambda}_1-\overline{\lambda}_2-\overline{\lambda}_3+\overline{\lambda}_4}{4}\ ,
\end{align}
which are the probabilities of the final Pauli channel.

Let us now generalise the notion of the age of a link.
Let $\left(t_1, t_2, \ldots, t_N\right)$ be the sequence of integers corresponding to the timesteps the $n$'th qubit was decohering for. The final Pauli channel is then given by 

\begin{align}
\mathcal{N}_1^{\circ t_1}\circ \mathcal{N}_2^{\circ t_2}\circ\ldots \mathcal{N}_N^{\circ t_N} \ .
\end{align}
Using that the fidelity of the Choi state is given by $\overline{p}_1$ and that $\overline{\lambda}_1 = 1$, we find that the fidelity is given by

\begin{align}
&~\frac{1+\overline{\lambda}_2+\overline{\lambda}_3+\overline{\lambda}_4}{4}\nonumber\\
=&~\frac{1+\left(\prod_{i=1}^N\left(\lambda_{i,2}\right)^{t_i} \right)+\left(\prod_{i=1}^N\left(\lambda_{i,3}\right)^{t_i} \right)+\left(\prod_{i=1}^N\left(\lambda_{i,4}\right)^{t_i} \right)}{4}\nonumber\\
=&~\frac{1+\sum\limits_{j=2}^{4}\exp(-\sum\limits_{i=1}^N{c_{i, j} t_i})}{4}\nonumber , 
\end{align}
where we set $\exp(-c_{i, j})= \lambda_{i, j}$. We thus find that the $\sum_{i=1}^{N}c_{i, j} t_i$ expressions can be interpreted as generalised \emph{age parameters} of the links. Note that in the homogeneous and depolarising noise setting all $\lambda$ parameters are equal, and one recovers the expression found in~\cite{haldar_reducing_2024}.

This is an alternative approach to other methods that used the maximum age of the parent links as a new link's age~\cite{inesta_optimal_2023, haldar_fast_2024}.

\section{MDP formulation}\label{app:mdp}

In this section we give a more precise description of the Markov decision process described in the main text. 

\paragraph{Action space:} 
This action space consists of actions $a$ of the form 

\begin{equation}
    a = (a_0, a_1, ..., a_{n-2}),
\end{equation}
where $a_i \in \{0,1\}$.

For the link generation round, when $a_i=1/0$, a link generation instruction is/is not sent to the $i\mhyphen\text{th}$ segment of the repeater chain respectively.

Similarly for the swap actions, when $a_i=1/0$, a swap instruction is/is not sent to the $i\mhyphen\text{th}$ node in the chain. 
Since node $0$ is an end-node and can only have one link, no swaps will be attempted at node $0$ regardless of the value. 

\paragraph{Observation space:}
Since the performed actions and the results uniquely determine the state, we use it directly as the observation for the agent. 
More precisely, for each node $i$ and qubit $j$, it keeps track of a list of tuples of the form 

\setlength{\arraycolsep}{0pt}
{
\small
\begin{equation}
\begin{bmatrix}
    (s(t), & r_{s(t-\Delta_{s})}, & e(t), & r_{e(t-\Delta_{\text{EG}})}) \\     (s(t-1), & r_{s(t-\Delta_{s}-1)}, & e(t-1), & r_{e(t-\Delta_{\text{EG}}-1)}) \\ 
     \vdots & \vdots & \vdots & \vdots \\
     (s(t-t_\text{cut}), & \; r_{s(t-\Delta_{s}-t_\text{cut})}, & \; e(t-t_\text{cut}), & \; r_{e(t-\Delta_{\text{EG}}-t_\text{cut}})
\nonumber
\end{bmatrix}_{i,j}.
\end{equation}
}

Here, $s(t) \in \{0,1\}$ represents whether a swap action has been sent out by the agent to node $i$ at time $t$ or not. 
 The result of the swap action that is sent out at time step $t-\Delta_{s}$ is denoted as $r_{s(t-\Delta_{s})}$, where  $\Delta_{s} = \Delta_{\text{swap}}(i,k) + \Delta_{\text{swap-result}}(i,k)$ is the corresponding delay.
Similarly, $e(t) \in \{0,1\}$ represents whether an entanglement generation action has been sent out to qubit $j$ in node $i$ at time step $t$ or not. 
The corresponding result that is received back at time step $t$ is denoted as $r_{e(t-\Delta_{\text{EG}})}$, where $\Delta_{\text{EG}} = 2  \Delta(i,k)$ if the index of the qubit $j=1$ and  $\Delta_{\text{EG}} = 2  \Delta(i-1,k)$ if the index of the qubit $j=0$. 
Note that the delays $\Delta_{s}$ and $\Delta_{\text{EG}}$ depend on the position $k$ of the global agent. 
When there is no result (i.e.~due to no action taking place at a particular time step) the value of the result is set to $-1$, and otherwise it is set to $1$/$0$ for success/fail, respectively.

\paragraph{Reward function:} The rewards are chosen so that the delivery time is minimised when the reward is maximised. 
For every time step that is not in the terminal state of the MDP, the reward function evaluates to $r = -1$. 
When the terminal state has been reached, the reward function evaluates to $r = 0$.
A state is a terminal state when end-to-end entanglement has been reached and held for $2 \text{max}(k, n-1-k)$ rounds. 
This is to ensure that the agent has had enough time to receive all of the relevant results.   

\paragraph{Environment dynamics:} 
In each time step, depending on whether it corresponds to the swap or link generation round, the corresponding values of $s(t)$ or $e(t)$ are being updated depending on which actions are sent out. 
Additionally, if it is the swap round, only the swap results are updated and if it is the link generation round only the link generation results are updated.

\section{Adapted swap-asap policies}\label{app:swap_asap}

For comparison reasons, the swap-asap policies presented in more detail will follow a similar structure to the Markov decision process on which the reinforcement learning policy is based. 
This means that in each time step, there are two rounds. That is, the first round is reserved for swap actions only and the second round is for entanglement generation actions only.

\subsection{Instantaneous swap-asap}

For the instantaneous swap-asap, after each round, we already assume that the policy can see the entire state of the quantum network. 
In each even round, it will perform a swap on all of the nodes where two links are present. 
In all of the odd rounds, it attempts entanglement generation on all segments on which both qubits are free, i.e.~the qubits on either side of the segment are not already involved in another link. 

\subsection{Wait-for-broadcast (WB) swap-asap}

The WB swap-asap with an agent located at node $k$ waits for 
$\Delta_{\text{EG}}$ time steps after sending out entanglement generation actions and it waits for $\Delta_{\text{swap}} + \Delta_{\text{swap-result}}$ time steps after sending out swap actions. 
This is to ensure that each time after sending out actions, enough time has passed such that all results have been collected back. 
If the previous time swap actions have been sent out, then in the current time step, it will choose to send out entanglement generation actions and vice versa. 
In each time step where it sends outs actions, the actions are chosen in the same way as the instantaneous swap-asap policy.  
It sends out entanglement generation instructions to all segments of which both qubits are free and it sends out swap instructions to all nodes with two links.
Note that as no actions are sent out when the agent is waiting for the results, we can effectively simulate the instantaneous swap-asap policy and multiply its delivery time by $\Delta_{\text{EG}} + \Delta_{\text{swap}} + \Delta_{\text{swap-result}}$, reducing the computation time.

\subsection{Predictive swap-asap}

We also introduce a version of the swap-asap policy which does not wait for global information but also does not have instantaneous access to the quantum network state. 
It still experiences classical communication effects, but instead of waiting, it chooses to predict the result of the action according to the success probabilities and perform its next action based on the predicted result. 
More specifically, each node attempts entanglement generation whenever a segment is predicted to be free from the node's point of view and it performs a swap when it thinks it has two nodes. 
The initial state (i.e.~the fully unlinked state) is always the same, and thus known.
The predictions are made randomly according to the success probabilities of each action. 
For a node $i$, it predicts that entanglement generation succeeds $p_e$ amount of the times. 
Once the node has predicted that entanglement generation has succeeded on both sides, it attempts a swap accordingly, which succeeds $p_s$ amount of the times. 
Only once the nodes have predicted that the quantum network state is end-to-end entangled do the nodes wait for one round of end-to-end communication, which takes $n-1$ times steps, to verify this.
Because this policy does not need the actual result of the actions, only the predicted result, to decide what to do next, we can let each node act locally, without waiting for communication from other nodes. In this case, the only waiting time comes from performing the entanglement generation action.
It should be noted that as each node acts and makes predictions locally without waiting for communication, the predictions between various nodes could differ. 
This makes it ambiguous when end-to-end entanglement has been predicted by the nodes. Additionally, node $i$ can for example predict that entanglement generation has succeeded with node $i+1$ whereas node $i+1$ has predicted that it failed. 
Node $i+1$ would then attempt entanglement generation again, but this cannot be done without node $i$ also sending a photon. 
To circumvent this, we require that all of the nodes start with the same random seed and make all of the same predictions for all of the actions and results in the network.
Each node will thus know which actions are performed by which other nodes and whether they have been predicted to succeed or not. 
Each node will thus also predict end-to-end entanglement at the same time.
Whether this prediction turns out to be correct will then be verified by one round of end-to-end communication.

\section{Numerical simulations}\label{app:num_sim}

The numerical simulations in this manuscript are based on two major components: the simulation of the environment and the reinforcement framework acting within that environment.

The most important part of the environment comprises of a quantum network simulator.
Since the simulation of the quantum network is only dealing with the (non-) existence of Bell pairs, we do not have to simulate the full statevector.
Instead, we can keep track of the current state of the network by tracking the success/failure of entanglement generation and swap operations.
We only keep track of the existing links in the linear network and their respective age.
If a link's age surpasses the cut-off, it is automatically deleted.

Due to classical communication delays, the agent does not directly observe the latest state of the network.
Instead, it might see a delayed version of the network's status, depending on the agent's position.
To construct the observation of the agent, we thus need to create the appropriate delayed history of the network.
Given the timestamp of each action and the position of the agent, we can generate a delayed history of the network which is passed to the agent.

The environment is written with the OpenAI gymnasium package~\cite{website_gymnasium}, which is compatible with stable-baselines3~\cite{raffin_stable-baselines3_2021} training algorithms. 
For the training, we have used the PPO algorithm~\cite{raffin_stable-baselines3_2021}. 
The training parameters are chosen as their default values, except the entropy coefficient for the loss calculation which was chosen as $0.001$. 
During training, we periodically save the agent at fixed checkpoints. 
From these checkpoints, we pick the policy which gives the highest rewards and use those for the evaluation of the policies. 

\section{Heat maps for various success probabilities}\label{app:heatmaps}

\begin{figure}[h]
    \centering
    \includegraphics[width=\columnwidth]{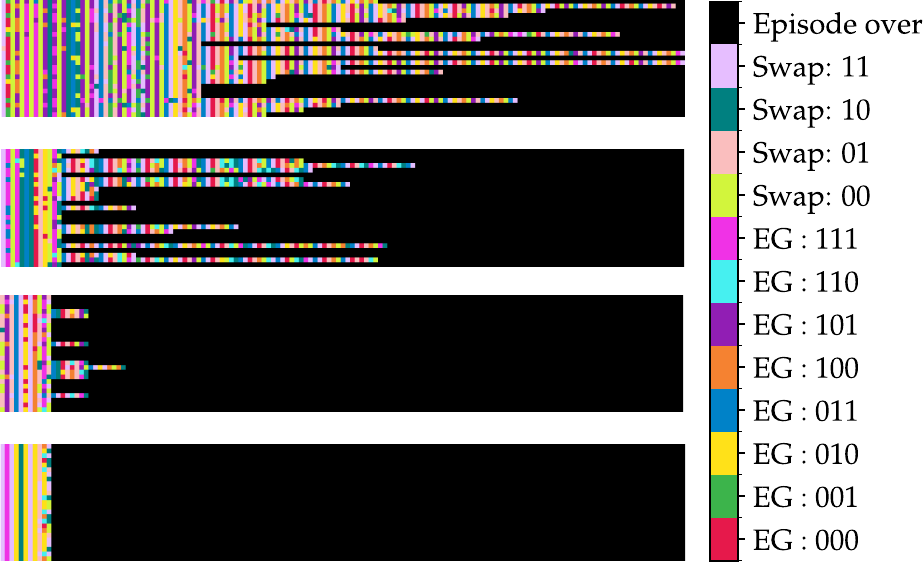}
    \caption{From top to bottom, the heat maps are ordered by success probabilities as: $p_s, p_e = 0.7$, $p_s, p_e = 0.8$, $p_s, p_e = 0.9$, $p_s, p_e = 1$}
    \label{fig:policy_action_plots}
\end{figure}

Here we show with Fig.~\ref{fig:policy_action_plots} that also for non-unit success probabilities, the final policies after training display strong patterns similar to those of the panel at the bottom right of Fig. \ref{fig:policy_interpretation}.
The panels of Fig. \ref{fig:policy_action_plots}  from the top to bottom are ordered by increasing success probabilities. 
The vertical axis corresponds to various simulation episodes and the horizontal axis corresponds to different rounds in an episode. 
Each pixel in the plot thus corresponds to an action taken at a specific round for a specific episode. 
We observe that as the success probabilities increase, the episodes on average get shorter, as expected. 
In the bottom most panel, each episode consists of $11$ rounds, which is the same as $5$ time steps, which is thus an optimal strategy for the setting where the agent is located at $2$ on a network of size $4$.
When considering the top most panel for example, we see that across different episodes with varying lengths, the same actions are often taken at the same time step. 
This suggest that even away from unit success probabilities, a reasonably good strategy often performs the same action in each round, regardless of the underlying state of the quantum network. 

\end{document}